\documentclass{elsart}
\usepackage{natbib}
\usepackage{graphicx}

\begin{document}

\runauthor{Antonio Capone and Giorgio Riccobene}

\begin{frontmatter}

\title{Measurements of light transmission in deep Sea with
the {\it AC9} trasmissometer}

 \author[UniRoma1,INFNRoma1]{A. Capone}
 \author[OGS]{T. Digaetano},
 \author[INFNCT]{A. Grimaldi},
 \author[UniINFNCA]{R. Habel},
 \author[INFNCT]{D. Lo Presti},
 \author[INFNLNS,UniCT]{E. Migneco},
 \author[UniRoma1]{R. Masullo},
 \author[OGS]{F. Moro},
 \author[INFNRoma1]{M. Petruccetti},
 \author[UniCT]{C. Petta},
 \author[INFNLNS]{P. Piattelli},
 \author[INFNCT]{N. Randazzo},
 \author[INFNLNS]{G. Riccobene \corauthref{ca:fax}},
 \ead{riccobene@lns.infn.it},
 \author[INFNRoma1]{E. Salusti},
 \author[INFNLNS]{P. Sapienza},
 \author[INFNLNS]{M. Sedita},
 \author[INFNLNF]{L. Trasatti} and
 \author[OGS]{L. Ursella}.

 \corauth[ca:fax]{Fax: +39 095 542 271}

\address[UniRoma1]{Dipartimento di Fisica Universit\'a La Sapienza, P.le A. Moro 2, 00185, Roma, Italy}
\address[INFNRoma1]{INFN Sezione Roma-1, P.le A. Moro 2, 00185, Roma, Italy}
\address[OGS]{Dipartimento di Oceanografia Fisica, Osservatorio Geofisico Sperimentale, Borgo Grotta Gigante 42C, 34016, Sgonico
(TS),Italy}
\address[INFNCT]{INFN Sezione Catania, Corso Italia 57, 95129, Catania, Italy}
\address[UniINFNCA]{Dipartimento di Fisica Universit\'a di Cagliari and Seione INFN Cagliari, 09042, Monserrato (CA),
Italy}
\address[INFNLNS]{Laboratori Nazionali del Sud INFN, Via S.Sofia 44, 95123, Catania, Italy}

\address[UniCT]{Dipartimento di Fisica e Astronomia Universit\'a di Catania}
\address[INFNLNF]{Laboratori Nazionali di Frascati INFN, Via Enrico Fermi 40, 00044, Frascati (RM), Italy}



\begin{abstract}
The NEMO Collaboration aims to construct an underwater
\v{C}erenkov detector in the Mediterranean Sea, able to act as a
neutrino telescope. One of the main tasks of this project, which
implies difficult technological challenges, is the selection of an
adequate marine site. In this framework the knowledge of light
transmission properties in deep seawater is extremely important.
The collaboration has measured optical properties in several
marine sites near the Italian coasts, at depths $>$3000 m, using a
set-up based on a {\it AC9}, a commercial trasmissometer,
manufactured by {\it WETLabs}. The results obtained for the two
sites reported in this paper ({\it Alicudi} and {\it Ustica}),
show that deep seawater optical properties are comparable to those
of the clearest waters. \vspace{1pc}
\end{abstract}

\end{frontmatter}

\section{Overview}
The observation of Ultra High Energy Cosmic Rays (UHECR) with
energy higher than 10$^{20}$ eV has attracted the attention of the
astrophysics and particle-physics community on the most energetic
phenomena taking place in the Universe. It is supposed that such
energetic particles are accelerated in extra-galactic sources.

Gamma ray sources with energy up to tens of TeV have also been
observed. If high energy photons are generated through the
production and decay of neutral pions, it is reasonable to expect,
from the same sources, an associated flux of high energy
neutrinos, generated through the production and decay of charged
pions. Along their journey in the universe, most part of the
electromagnetic and hadronic emission is deflected or absorbed by
the electromagnetic background and by the intergalactic and
interstellar matter. Neutrinos, on the contrary, are not
significantly absorbed by the intergalactic medium and are not
deflected by the intergalactic magnetic fields. Already in 1960
Markov \cite{markov1960,markov1961} proposed to use seawater as a
huge target to detect UHE neutrinos, looking at their charged
current weak interactions. The outgoing lepton generates, along
its path in seawater, \v{C}erenkov light that can be detected by a
lattice of optical sensors. The reconstruction of the muon track,
and thus of the neutrino direction, offers the possibility to
identify the neutrino sources opening the new exciting field of
neutrino astronomy. The observation of high energy neutrino fluxes
expected from astrophysical sources requires a detector with an
effective area close to 10$^6$ m$^2$ instrumented along a distance
comparable to the range in water ($\sim$ km) of the high energy
muons ($\sim$ 1000 TeV). The identified neutrino sources
identified could be catalogued in the sky map and eventually
compared with the known gamma sources. The construction of a
detector of such dimensions, usually called a km$^3$ Neutrino
Telescope, is one of the main challenges of astroparticle physics
today. The Mediterranean Sea offers optimal conditions, on a
worldwide scale, to locate an underwater neutrino telescope. The
choice of the km$^3$ scale neutrino telescope location is such an
important task that careful studies of candidate sites must be
carried out in order to identify the most suitable one. Along the
Italian coasts several sites exist, at depth $3300 \div 3500$ m,
that are potentially interesting to host an undersea neutrino
telescope. In these sites we have studied deep seawater optical
properties (absorption and attenuation) and environmental
properties: water temperature and salinity, biological activity,
water currents, sedimentation. In this paper we report light
transmission measurements carried out in two sites named {\it
Ustica} (during November 1999) and {\it Alicudi} (on December
1999), in the Southern Tyrrhenian Sea, located at:

\begin{itemize}
 \item {\bf $39^{\circ} 05'$ N $13^{\circ} 20'$ E}, North-Est of
Ustica island;
 \item {\bf $39^{\circ} 05'$ N $14^{\circ} 20'$ E}, North of
Alicudi island.
\end{itemize}

\section{Optical properties of deep sea}

Water transparency to electromagnetic radiation can be
characterized by means of quantitative parameters: the absorption
length $L_a$ and the scattering length $L_b$. Each length
represents the path after which a photon beam of intensity $I_0$
at wavelength $\lambda$, travelling along the emission direction,
is reduced to $1/e$ by absorption or diffusion phenomena. These
quantities can be directly derived by the simple relation:

\begin{equation}
I(x,\lambda) = I_0 (\lambda)  e^ {- ~ \frac{x}{L(\lambda)} },
\label{eq:lambertlaw}
\end{equation}

\noindent where $x$ is the optical path traversed by the beam and
$I_0$ the source intensity. In literature absorption ($a =1/L_a$)
and scattering ($b =1/L_b$) coefficients are extensively used to
characterize the light transmission in matter as well as the
attenuation coefficient ($c$) defined as:

\begin{equation}
c (\lambda) = a(\lambda) + b(\lambda). \label{eq:attenuation}
\end{equation}

The main cause of light absorption in water is excitation of
vibrational states of the water molecule by photons
\cite{duntley1963,Warren1984,Braun1993}: due to such process the
photon energy is entirely deposited in the traversed medium.
Scattering refers to processes in which the direction of the
photon is changed without any other alteration. Scattering
phenomena in which the photon wavelength changes (e.g. Raman
effect) happen less frequently. Scattering can take place either
on molecules (Rayleigh scattering) or on dissolved particulate
(Mie scattering).

\begin{figure}[bt]
\label{fig:waterspectrumduntley}
\centerline{\includegraphics[width=11cm]{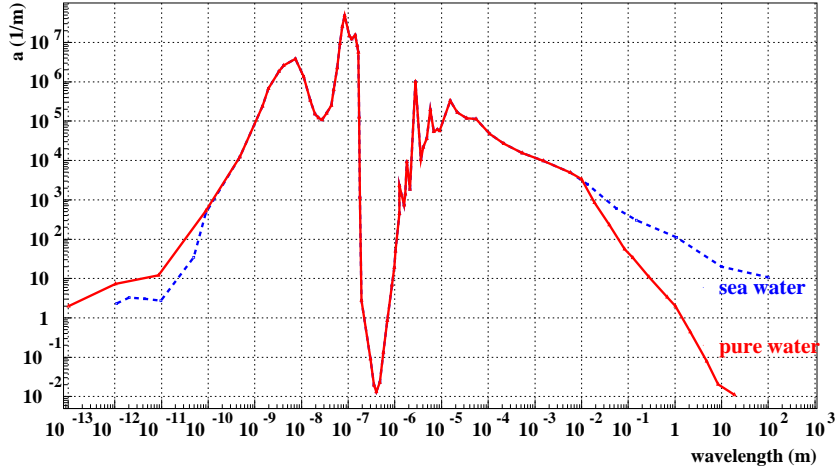}}
\caption{ Absorption coefficient of electromagnetic waves for pure
and seawater as function of wavelength. Data taken from Mobley
\cite{Mobley1994}.}
\end{figure}

In pure water, light absorption and scattering are strongly
wavelength dependent. In particular light transmission in pure
water is extremely favored in the range 350 $\div$ 550 nm,
overlapping the region in which PMTs usually reach the highest
quantum efficiency. In the visible region of the electromagnetic
spectrum light absorption steeply decreases as a function of
wavelength  and reaches its minimum at about 420 nm (see figure
\ref{fig:waterspectrumduntley}). This is the reason why seawater
has a blue-green color.

The optical properties of natural seawater are functions of water
salinity, water temperature and of the concentration of dissolved
organic and inorganic matter. Light absorption and diffusion in
water as a function of salinity and temperature have been
extensively studied \cite{pegau97}. It has been noticed that, for
$\lambda\geq 400$ nm, the dependence of scattering coefficient on
temperature and salinity is negligible while the variation of the
absorption coefficient is significant, in particular at $\lambda >
710$ nm (for details see \S \ref{sec:dataanalysis}). The seawater
diffusion and absorption coefficients can be parameterized as the
sum of a term due to optically pure water (i.e. water without
dissolved particulate) at defined conditions of temperature and
salinity ($a_{W}^{T,S},b_{W}^{T,S}$), and a term that accounts for
interaction of light with particulate ($a_p,b_p$):

\begin{equation}
a_{SW} (\lambda)= a_{W}^{T,S} (\lambda) + a_p (\lambda),
\label{eq:totalabsorption}
\end{equation}

\begin{equation}
b_{SW} (\lambda)= b_{W}^{T,S} (\lambda) + b_p (\lambda).
\label{eq:totaldiffusion}
\end{equation}

Optical measurements of deep seawater have shown that the presence
of particulate has a negligible effect on light absorption but it
enlarges the light diffusion coefficient.
Since water temperature
and salinity and particulate concentration may vary significantly
in different marine sites it is extremely important to measure
optical parameters {\it in situ}.

\section{The AC9 trasmissometer}

We performed attenuation and absorption measurements of light in
deep seawater by means of a set-up based on a trasmissometer: {the
\it AC9} manufactured by {\it WETLabs} \cite{Wetlabsmanual}. The
device compactness (68 cm height $\times$ 10.2 cm diameter) and
its pressure resistance (it can operate down to 6000 m depth) are
excellent for our purposes. The {\it AC9} performs attenuation and
absorption measurements independently using two different light
paths and spanning the light spectrum over nine different
wavelengths (412, 440, 488, 510, 532, 555, 650, 676, 715 nm). In
our measurements we obtain an accuracy in $a(\lambda)$ and
$c(\lambda)$ of about $1.5 \cdot 10^{-3}$ m$^{-1}$.

The {\it AC9} attenuation and absorption measurement technique is
based on the Lambert's law of collimated beams (see equation
\ref{eq:lambertlaw}) where $x$ is the beam path-length, $I_0
(\lambda)$ is the intensity of the collimated primary beam, at a
given wavelength $\lambda$, and $I(x,\lambda)=I_{a,c}(x,\lambda)$
is the beam intensity measured at distance $x$, as a result of
absorption or attenuation effect respectively.

In order to produce collimated monochromatic light beams, the
instrument is equipped with an incandescence lamp and a set of
collimators and nine monochromatic ($\Delta \lambda \sim $ 10 nm)
filters . Two different beams are available at the same time for
independent measurements of attenuation and absorption. Each beam
is split in two parts by a mirror: the reflected one reaches a
reference silicon photon detector. The refracted one crosses a
quartz window and enters inside a 25 cm long pipe. During deep sea
measurements seawater fills the pipes ({\it flow tubes}). The flow
tube used for attenuation measurements has a black inner surface
in order to absorb all photons scattered by seawater. A collimated
silicon photon detector (angular acceptance $\sim 0.7^{\circ}$) is
placed at the end of the path, along the source axis. Thanks to
this strongly collimated layout the end-path detector receives
only photons which have not interacted (neither absorbed nor
scattered). The reference detector measures the source intensity
$I_0(\lambda)$, the end-path detector measures the attenuated beam
intensity $I_c(x,\lambda)$, $x$ is the known beam path in water
(0.25 m). The attenuation coefficient is therefore calculated as:

\begin{equation}
c (\lambda) = \frac{1}{x} \ln \frac{I_0 (\lambda)}{I_c
(x,\lambda)}.
\label{eq:inverselambertlaw}
\end{equation}

In the absorption channel, the inner surface of the {\it
absorption flow tube} behaves like a cylindrical mirror. The light
scattered by seawater is reflected and redirected towards a wide
angular acceptance silicon photon detector. In first approximation
all scattered photons are detected and the ratio between the
intensities $I_0(\lambda)$ and $I_a(x,\lambda)$ is only a function
of the seawater absorption coefficient $a(\lambda)$.

Using {\it AC9} data, the scattering coefficient can be calculated
by subtracting the absorption value from the attenuation value at
each given wavelength (see equation \ref{eq:attenuation}).

\section{AC9 measurements principles}

In the interval of wavelength interesting for a \v{C}erenkov
neutrino telescope, $\lambda = 350 \div 550$ nm, the expected
values of absorption and attenuation coefficients in deep seawater
are $\sim 10^{-2}$ m$^{-1}$, very close to the pure water ones.
This implies that the instrument should have sensitivity and
accuracy of the order of $1 \div 2 \cdot 10^{-3}$ m$^{-1}$. The
calibration of the instrument plays the most important role in
determining the accuracy in measurements. In the above defined
wavelength range, pure water optical properties have been
extensively measured, therefore pure water can be assumed as a
reference medium \cite{Kou1993,Pope93}.

Instrumental effects - such as the status of optical windows, of
the electronics, etc.- can also be studied filling the flow tubes
with a medium with a negligible light absorption and attenuation
(e.g. dry air or $N_2$).

The instrument calibration can be, thus, performed and tested any
time filling the flow tubes with a medium with known optical
properties: either pure water ({\it pure water calibration}) or
$N_2$ ({\it air calibration}). Filling the flow tubes with pure
water we measure, for example in the absorption channel, the
values:

\begin{equation}
a (\lambda) = a_{I}(\lambda) + a_{PW}(\lambda)
\label{eq:calibawater}
\end{equation}

\noindent and, in the case of $N_2$,

\begin{equation}
a (\lambda) = a_{I}(\lambda) + a_{N_2}(\lambda).
\label{eq:caliban2}
\end{equation}

The extra-term $a_{I}(\lambda)$ takes into account the light
absorption in the instrument optics (that is function of the
status of quartz windows and mirror surfaces) and all other
instrumental effects (see \S \ref{sec:dataanalysis}). This means
that this term can vary with time and can be a function of the
internal electronics temperature. The same argument is valid for
the attenuation channel.

With pure water inside the flow tubes the measurement of {\it AC9}
can be set equal to the known values of $a_{PW}(\lambda)$ and
$c_{PW}(\lambda)$. The result of the {\it water calibration}
procedure is a set of 18 calibration constants (for the nine
absorption and attenuation channels) that represent the working
status of {\it AC9}. The {\it AC9} internal software subtracts
these coefficients to the actual reading of the instrument such
that each AC9 output value, when reference water fills the flow
tubes, should be equal to zero.

Filling the flow-tubes with sea water we measure:

\begin{equation}
a(\lambda) = a_{I}(\lambda) + a_{SW}(\lambda)
\label{eq:measureabsorption}
\end{equation}

\begin{equation}
c(\lambda) = c_{I}(\lambda) + c_{SW}(\lambda).
\label{eq:measureattenuation}
\end{equation}

\noindent Then the values of AC9 output corresponding to the case
of deep sea water filling the flow tubes are:

\begin{equation}
\Delta a_{SW}(\lambda) = a_{SW}(\lambda)- a_{PW}(\lambda),
\label{eq:delta_a_def}
\end{equation}

\begin{equation}
\Delta c_{SW}(\lambda) = c_{SW}(\lambda)- c_{PW}(\lambda).
\label{eq:delta_c_def}
\end{equation}

\section{AC9 calibration procedure}
\label{sec:ac9calibration}

The {\it AC9} manufacturer ({\it WETLabs}) provides the instrument
calibration performed with de-ionized and de-gassed pure water, at
given temperature ($25 ^{\circ}$C), as referenced in
\cite{Kou1993,Pope93}. The optical properties of this medium, at
the nine wavelengths relevant for the {\it AC9}, are listed in
table \ref{tab:purewater}. {\it Wetlabs} provides also the results
of the instrument calibration performed with dry air. The set of
constants that relate the {\it water calibration} values to the
{\it air calibration} ones are provided by {\it WETLabs}.

In principle, in order to test, from time to time, the validity of
the used set of {\it water calibration} constants, the user should
check the {\it AC9} response after filling the flow tubes with
pure water in {\it reference} conditions. However, since pure
water is not easily available during cruises, we check the
calibration of the {\it AC9} in dry air testing the validity of
air calibration constants after having filled the flow tubes with
high purity grade ${N_2}$. We extensively perform these operations
during naval campaigns before every deployment.

\begin{table}[tb]
\caption{Absorption and attenuation coefficients (in units of
10${-3}$) of pure water at $T=25^\circ$ referenced in the {\it
AC9} manual \cite{Wetlabsmanual} and \cite{Pope97}}
\label{tab:purewater}
\begin{center}
\begin{tabular}{rrrrrrrrrr} \hline
  $\lambda$ (nm)          & 412 &  440 &  488 &  510 &  532 &  555 &   650 &   676 & 715\\
  $a$ (10$^{-3}$ m$^{-1}$) & 5.4 &  8.3 & 17.7 & 38.2 & 51.6 & 69.0 & 359.4 & 441.6 & 1049.2 \\
  $c$ (10$^{-3}$ m$^{-1}$) & 9.7 & 11.9 & 20.0 & 40.2 & 53.3 & 70.4 & 360.1 & 442.2 & 1049.7 \\
\hline
\end{tabular}
\end{center}
\end{table}

\begin{figure}[hbt]
\centerline{\includegraphics[width=11 cm]{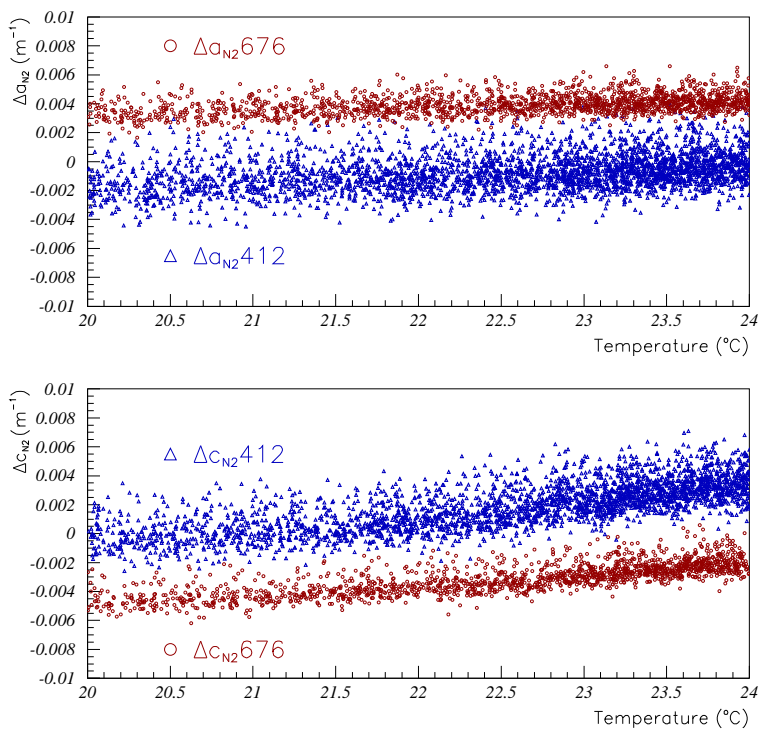}}
\caption{Absorption $\Delta a_{N_2}(\lambda)$ and attenuation
$\Delta c_{N_2}(\lambda)$ raw values for $\lambda=412,676$ nm,
measured, as function of the {\it AC9} internal temperature,
during a dry air calibration before the first measurement in
Alicudi site.}
\label{fig:aircoefficients}
\end{figure}

We have accurately studied the dependence of {\it air calibration}
constants as a function of {\it AC9} internal temperature. We
noticed that during measurement in the Mediterranean Sea, where at
depth$>1500$ m the water temperature is $\sim 13 \div 14
^{\circ}$C, the {\it AC9} internal temperature stabilizes at
$T_{AC9}\sim 22.5^{\circ}$C. To reduce the systematic error in the
knowledge of the instrument calibration constants, during checks
of {\it air calibration} we keep the internal {\it AC9}
temperature at $\sim22.5^{\circ}$C by means of a refrigerator.
Figure \ref{fig:aircoefficients} shows some {\it AC9} measured raw
values $\Delta a_{N_2}(\lambda), \Delta c_{N_2} (\lambda)$
(analogous to $\Delta a_{SW}(\lambda)$ and $\Delta
c_{SW}(\lambda)$ as defined in equations \ref{eq:delta_a_def} and
\ref{eq:delta_c_def}), as a function of $T_{AC9}$, obtained
checking the {\it air calibration} just before the first
deployment in {\it Alicudi} site. In the range $22^{\circ}$C$ <
T_{AC9} < 23^{\circ}$C the $\Delta a_{N_2}(\lambda)$, $\Delta
c_{N_2} (\lambda)$ average values are close to zero and RMS are of
the order of $1.5 \cdot 10^{-3}$ m$^{-1}$. These average values
(that we call $a_{N_2}^{off}(\lambda)$ and
$c_{N_2}^{off}(\lambda)$ ) are used as {\it offsets} and
subtracted during off-line data analysis, as described in the
following sections.

\section{Deep Sea Setup}
\label{sec:setup}

\begin{figure}[hbt]
\centerline{\includegraphics[width=6.5cm]{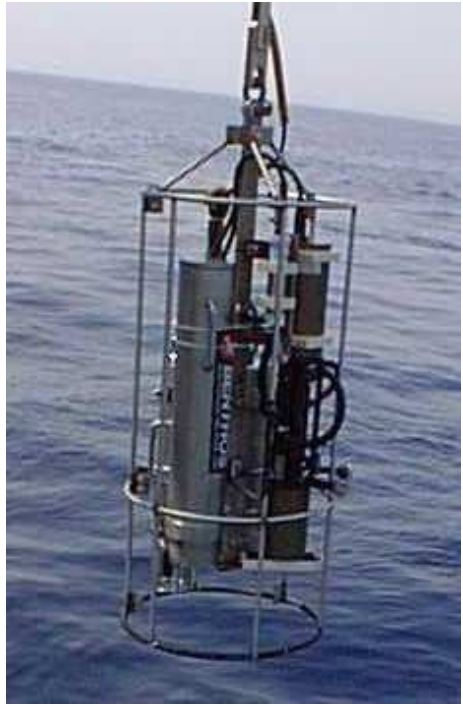}}
\caption{Deployment of AC9 deep-sea setup.} \label{fig:photocage}
\end{figure}

During deep sea measurements the {\it AC9} is connected to a
standard oceanographic CTD (Conductivity Temperature Depth) probe,
the {\it Ocean ~ MK-317} manufactured by {\it IDRONAUT}. A pump is
used to ensure re-circulation of seawater inside the {\it AC9}
pipes. The {\it AC9} and the pump are powered by a 14 V battery
pack. In figure \ref{fig:photocage} we show the whole set-up
mounted on an AISI-316 stainless-steel cage before a deployment.

\begin{figure}[htb]
\centerline{\includegraphics[width=11cm]{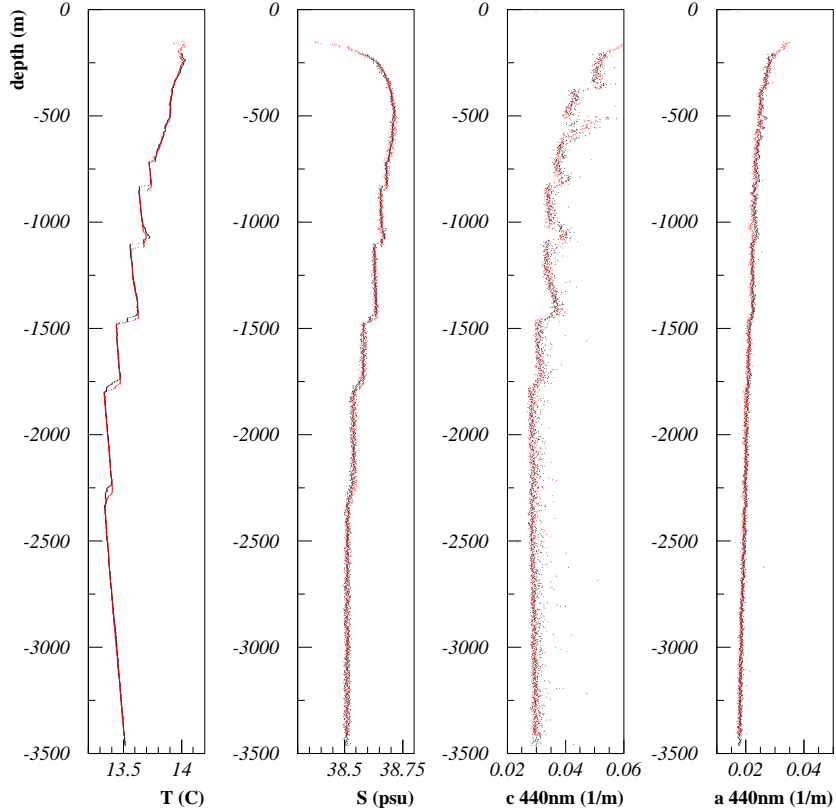}}
\caption{Temperature, salinity, attenuation and absorption
coefficients (at $\lambda=440$ nm) as a function of depth,
measured in the first (black) and in the second (red) deployment
in {\it Alicudi} site. The values measured in the two deployments
are nearly superimposed in the figure.}
\label{fig:SalTempacvsdepth}
\end{figure}

When the system is in operation, the {\it RS-232} stream of the
{\it AC9} data is converted into {\it FSK} stream by a modem card
placed inside the CTD. Data are sent to sea surface through an
electro-mechanical cable, that is also used to transmit power to
CTD ($\sim$1 A at 30 VDC). The data acquisition system permits
both the {\it AC9} data telemetry and the data storage on a PC
onboard the ship. The CTD data (depth, water salinity and
temperature) are recorded on a local memory. Both instruments also
record an absolute time information, which is used to couple the
{\it AC9} and the CTD data during off-line analysis. This
procedure allows to relate water optical properties to depth,
water salinity and temperature. The CTD-AC9 acquisition program
gives about six measurements per second, usually we deploy the
set-up at speed of 1 m/s. In figure \ref{fig:SalTempacvsdepth} we
show, as function of depth, water salinity and temperature
together with the final absolute values of absorption and
attenuation coefficient for $\lambda=440$ nm measured during the
first (black dots) and the second (red dots) deployments in
Alicudi site. As it appears in the figure, the layer composition
of Tyrrhenian Sea, well studied by oceanologists in terms of
salinity and temperature, is also indicated by the measurements of
water optical properties: the AC9 sensitivity permits to
distinguish layers of water where absorption and attenuation
coefficients vary for $\sim1\cdot 10^{-3}$ m$^{-1}$.

\section{Data Analysis}
\label{sec:dataanalysis}

In order to obtain the values of deep seawater absorption and
attenuation coefficients from the measured raw values we need to
apply few corrections. The first correction consists in removing
the set of calibration constants $a_{N_2}^{off}(\lambda),
c_{N_2}^{off}(\lambda)$ described in \S \ref{sec:ac9calibration}:

\begin{equation}
\Delta a'(\lambda)= \Delta a_{raw}(\lambda) -
a_{N_2}^{off}(\lambda),
\label{eq:absorptioncorrection1}
\end{equation}

\begin{equation}
\Delta c'(\lambda)= \Delta c_{raw}(\lambda) -
c_{N_2}^{off}(\lambda). \label{eq:attenuationcorrection1}
\end{equation}

Further corrections for the attenuation channel could be needed to
take into account that the silicon photon detector in the
attenuation channel has a finite angular acceptance
($0.7^{\circ}$) and that the inner surface of the {\it attenuation
flow tube} does not behave as a perfect absorber. These two
corrections have been evaluated to be much lower than $1.5 \cdot
10^{-3}$ m$^{-1}$ that we quote as systematic error associated to
the result. Therefore $\Delta c_{corr}(\lambda)= \Delta
c'(\lambda)$.

The absolute values of the light attenuation coefficients in
seawater (as a function of depth) can be finally obtained
inverting equations \ref{eq:delta_c_def}.

Concerning the absorption channel, up to now, we considered that
the inner surface of the {\it flow tube} is perfectly reflecting.
This assumption is valid only in first approximation and a proper
correction has to be applied to the measured raw values.  If the
inner mirror is not perfectly reflecting, in presence of light
scattering, a fraction of the diffused photons do not reach the
end-path detector. We now illustrate how we have evaluated the
amount of this effect using {\it AC9} data collected for $\lambda
>650$ nm.

Photon diffusion in the absorption channel is also present when
the tube is filled with pure water. This implies that, with the
described calibration procedure, part of the effect is already
accounted for at the calibration time, i.e. the effect of photons
diffused at large angle by Rayleigh scattering on molecules.

The presence of particulate in deep seawater results as an
additional cause of absorption and diffusion processes; but, at
red and infrared wavelengths, the absorption due to the
particulate present in deep seawater is negligible
\cite{Twardowski1999}. It follows that the values $\Delta
a'(\lambda)$ at $\lambda= 676$ nm and $\lambda= 715$ nm, measured
by the {\it AC9} in deep seawater, allow us to evaluate the effect
of the not perfect reflectivity of the {\it absorption flow tube}
mirror.

Actually a value of $\Delta a'(\lambda) \neq 0$ for seawater has
to be expected because of the presence of salts and because deep
seawater temperature ($\sim 13 \div 14 ^\circ$C in the deep
Mediterranean Sea) is not equal to the calibration water
temperature ($25^\circ$C). The dependence of light absorption and
diffusion in water as a function of salinity and temperature has
been extensively studied \cite{pegau97}. It has been noticed that,
for $\lambda > 400$ nm, the dependence of $b_W^{T,S}(\lambda)$ on
temperature and salinity is negligible; on the contrary the
variation of the absorption coefficient can be expressed by the
equation:

\begin{equation}
\Delta a^{T,S}(\lambda)=[\Psi_T \cdot (T-T_{ref}) + \Psi_S \cdot
(S-S_{ref})]
\label{eq:salinitytemperature}
\end{equation}

\noindent where $T_{ref}=25^{\circ}$C, $S_{ref}=0$ practical
salinity units (p.s.u.), $T$ and $S$ are the actual values of
seawater. The constants $\Psi_T$ and $\Psi_S$ are known as a
function of the wavelength \cite{pegau97,Twardowski1999}: for
$\lambda=676$ nm the values are $\Psi_T(676)=1 \cdot 10^{-4}$
m$^{-1}$ $^{\circ}$C$^{-1}$ and $\Psi_S(676)=8\cdot10^{-5}$
m$^{-1}$ {p.s.u.}$^{-1}$. The slope of the temperature corrections
for $\lambda=715$ nm is much larger: $\Psi_T(715)=2.9 \cdot
10^{-3}$ m$^{-1}$ $^{\circ}$C$^{-1}$ (while $\Psi_S(715)= -8 \cdot
10^{-5}$ m$^{-1}$ {p.s.u.}$^{-1}$).

We evaluate the correction due to the internal mirror of the {\it
absorption flow tube} only at $\lambda=676$ nm since the
uncertainty on the temperature correction at this wavelength is
smaller.

The evaluation of the contribution to the value of $\Delta
a'(676)$ due to the photons diffused by particulate and not
reflected toward the end-path detector is determined by means of
the equation:

\begin{equation}
\Delta a^{mirror}(676)= \Delta a'(676) - \Delta a^{T,S}(676).
\label{eq:reflcorr}
\end{equation}

The measurement of $S_{SW}$ and $T_{SW}$, made by the CTD (see \S
\ref{sec:setup}), allows to evaluate the correction $\Delta
a^{mirror}(676)$ as a function of depth.

It has been suggested \cite{Zaneveld94} that the shape and
magnitude of the Mie volume scattering function, in first
approximation, can be considered almost independent on wavelength
for the interval of $\lambda$ in which the {\it AC9} operates: the
correction due to the {\it mirror} effect is, therefore,
independent on wavelength. It turns out that $\Delta
a^{mirror}(676)$ can be used to correct the measured values of
absorption coefficients for all wavelengths. Applying this
correction we obtain:

\begin{equation}
\Delta a_{corr}(\lambda)= \Delta a'(\lambda) - \Delta
a^{mirror}(676).
\label{eq:absorptioncorrection2}
\end{equation}

Finally, adding the pure water absorption ($a_{PW}$) and
attenuation ($c_{PW}$) coefficients to the obtained $\Delta
a_{corr}$ and $\Delta c_{corr}(\lambda)$, we evaluate (as a
function of depth) the seawater {\it inherent optical properties}:

\begin{equation}
a_{SW}(\lambda)= \Delta a_{corr}(\lambda) + a_{PW}(\lambda),
\label{eq:absorptioninsitu}
\end{equation}

\begin{equation}
c_{SW}(\lambda)= \Delta c_{corr}(\lambda) + c_{PW}(\lambda).
\label{eq:attenuationinsitu}
\end{equation}

\begin{figure}[htb]
\centerline{\includegraphics[width=11cm]{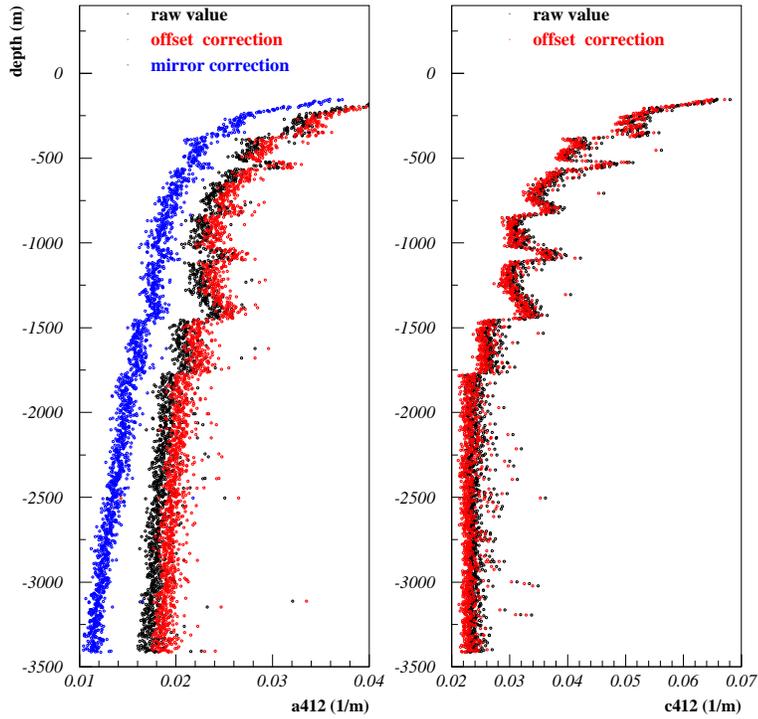}}
\caption{Raw values (black dots) of the absorption and attenuation
coefficients at 412 nm. The figure also shows the values obtained
after {\it offset correction} (red dots) and, only in the case of
absorption, the values after {\it mirror correction} (blue dots).}
\label{fig:analisiac}
\end{figure}

Figure \ref{fig:analisiac}, that refers to the second measurement
in Alicudi site, illustrates the analysis procedure for
$\lambda=412$ nm. As function of depth, we show the raw measured
values of the absorption and attenuation coefficients (black
dots), the values obtained applying the {\it offset} correction
(red dots) and the {\it mirror} correction (blue dots) for
$a_{SW}(412)$ and $c_{SW}(412)$.

The same analysis has been applied to the values measured at
$\lambda = 440$, 488, 510, 532, 650, 676 and 715 nm. The values of
$a_{SW}(\lambda)$ and $c_{SW}(\lambda)$ measured in the four
deployments carried out in {\it Alicudi} and {\it Ustica} show
very good agreement.

We do not show results for $\lambda=555$ nm due to a temporary
hardware problem happened to the interferometric filter during the
naval campaign.

In the following we will quantify the optical properties of deep
seawater averaging the absorption and attenuation coefficients in
the range of depth interesting for a km$^3$ neutrino detector: a
400 m wide interval, with its base $\sim 150$ m above the seabed.

\begin{figure}[htb]
\centerline{\includegraphics[width=11cm]{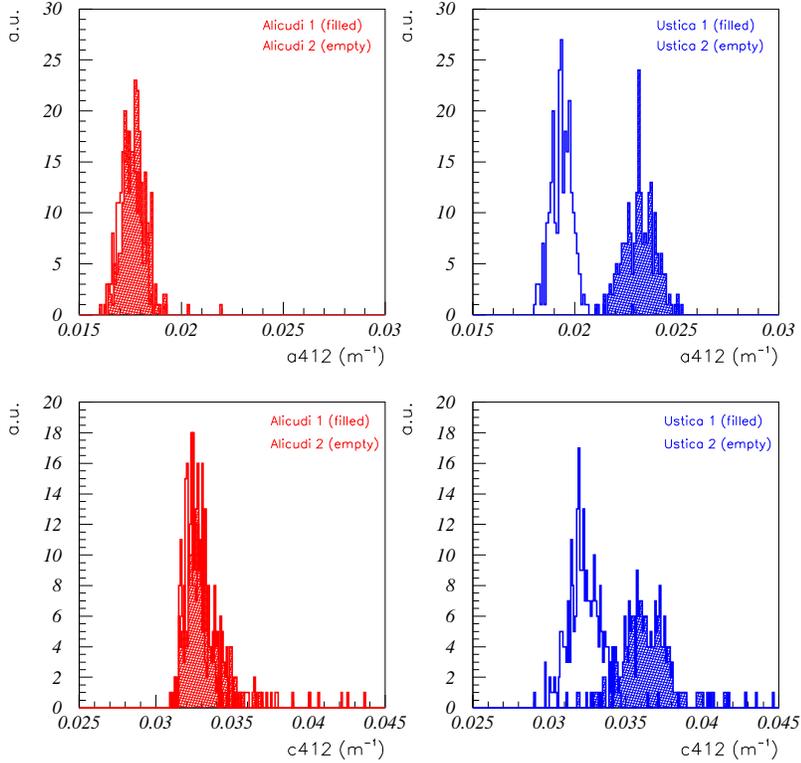}}
\caption{Distributions of the absorption and attenuation
coefficients, at $\lambda=412$ nm, measured in $Alicudi$ and
$Ustica$ in the depth interval $2850 \div 3250$ m. The results for
$a_{412}$ and $c_{412}$ for the two measurements in {\it Alicudi}
are in excellent agreement and appear nearly superimposed in the
figure.} \label{fig:AC412histos}
\end{figure}

In figure \ref{fig:AC412histos} we show, as an example, the
distribution of $a_{SW}(412)$ and $c_{SW}(412)$ values, averaged
for each meter, in the interval of depth $2850 \div 3250$ m,
related to the deployments in {\it Ustica} and {\it Alicudi}
(seabed depth $\sim 3400$ m for both sites).

In table \ref{tab:deltacorrected} we list the average values of
the distributions of $\Delta a_{corr}(\lambda)$ and $\Delta
c_{corr}(\lambda)$ in the same interval of depth.

\begin{table*}[tb]
\label{tab:deltacorrected} \caption{Values of $\Delta a_{corr}$
and $\Delta c_{corr}$ measured in the interval of depth 2850
$\div$ 3250 m in $Alicudi$ and $Ustica$ during November-December
1999. Negative values, when statistically significant, are due to
the dependence of water optical properties on salinity and
temperature.} \label{tab:measuredvalues}
\begin{center}
\begin{tabular}{rrrrr}
\hline
  coefficient & Alicudi-1& Alicudi-2 & Ustica-1 & Ustica-2 \\
        & $\cdot10^{-3}[m^{-1}]$ & $\cdot10^{-3}[m^{-1}]$ & $\cdot10^{-3}[m^{-1}]$ & $\cdot10^{-3}[m^{-1}]$ \\
\hline
  a412 & $12.3\pm0.6$ & $12.4\pm0.5$ & $17.8\pm0.8$ & $14.2\pm0.5$ \\
  c412 & $23.6\pm1.8$ & $23.3\pm1.7$ & $28.5\pm1.7$ & $22.8\pm2.0$ \\
  a440 & $10.1\pm0.5$ & $9.9\pm0.5$  & $13.2\pm0.6$ & $10.9\pm0.5$ \\
  c440 & $18.1\pm1.8$ & $17.8\pm1.7$ & $19.8\pm1.8$ & $17.7\pm1.9$ \\
  a488 & $4.4\pm0.4$  & $4.3\pm0.4$  & $5.5\pm0.5$  & $4.5\pm0.5$  \\
  c488 & $13.8\pm1.8$ & $13.7\pm1.7$ & $14.1\pm1.6$ & $14.2\pm1.6$ \\
  a510 & $-0.7\pm0.4$ & $-0.6\pm0.3$ & $0.4\pm0.5$  & $-0.2\pm0.5$ \\
  c510 & $0.6\pm1.9$  & $0.6\pm1.7$  & $0.4\pm1.6$  & $1.7\pm1.6$  \\
  a532 & $0.8\pm0.4$  & $1.2\pm0.3$  & $1.3\pm0.5$  & $1.2\pm0.4$  \\
  c532 & $-3.1\pm1.9$ & $-4.0\pm1.7$ & $-3.7\pm1.6$ & $-1.8\pm1.8$ \\
  a650 & $-3.1\pm0.4$ & $-2.5\pm0.5$ & $-2.6\pm0.4$ & $-1.8\pm0.4$ \\
  c650 & $15.7\pm1.9$ & $18.4\pm1.9$ & $16.2\pm1.5$ & $17.2\pm1.7$ \\
  a676 & $0\pm0.5$    & $0\pm0.5$    & $0\pm0.5$    & $0\pm0.6$ \\
  c676 & $3.7\pm1.8$  & $2.9\pm1.6$  & $5.6\pm1.1$  & $-0.9\pm1.6$ \\
  a715 & $-33.\pm1.0$  & $-32.\pm1.0$  & $-33.\pm1.0$  & $-33.\pm1.0$ \\
  c715 & $-33.\pm2.0$  & $-33.\pm1.6$ & $-33.\pm1.2$ & $-33.\pm1.5$ \\
\hline
\end{tabular}
\end{center}
\end{table*}

The statistical errors associated to these values are evaluated
from the RMS of the distributions (see, for example, figure
\ref{fig:AC412histos}). The systematic errors, mainly due to the
accuracy of the calibration check procedure, have been evaluated
to be equal to $1.5\cdot10^{-3}$. The absolute ($a_{SW}(\lambda),
c_{SW}(\lambda)$) values can be obtained adding the values of
attenuation and absorption of the {\it reference} water (see
equations \ref{eq:absorptioninsitu},\ref{eq:attenuationinsitu} and
table \ref{tab:purewater}).

\begin{figure}[htb]
\centerline{\includegraphics[width=11cm]{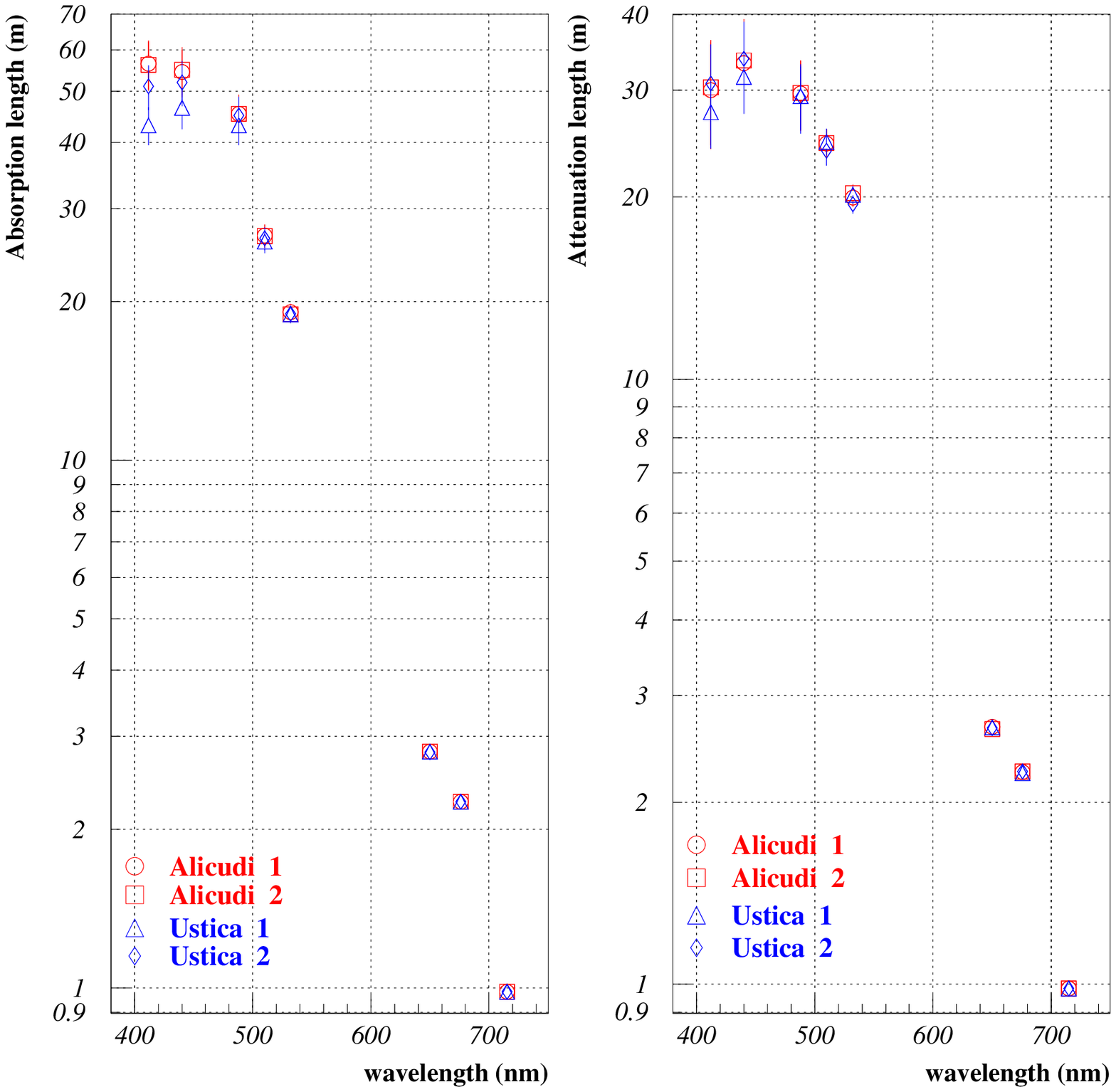}}
\caption{Absorption and attenuation lengths as function of
wavelength measured in $Alicudi$ and $Ustica$, in the depth
interval $2850 \div 3250$ m. The error bars include the systematic
error.} \label{fig:resultslengths}
\end{figure}

Finally, we present in figure \ref{fig:resultslengths} the values
of absorption and attenuation lengths, as a function of
wavelength, for the same interval of depth.

\section{Discussion}

The importance of measuring seawater optical properties {\it in
situ} has been discussed by several authors. The set-up we used
permits to evaluate seawater light absorption and attenuation
coefficient as a function of depth and wavelength (in the range
$412 \div 715$ nm). Accurate calibration checks allow us to obtain
an accuracy in the evaluation of $a(\lambda)$ and $c(\lambda)$ of
the order of $1.5 \cdot 10^{-3}$ m$^{-1}$.

The values of $a_{SW}(\lambda)$ measured in the depth interval of
interest in {\it Alicudi} and {\it Ustica} are very close to the
ones reported by Smith and Baker as {\it pure seawater} absorption
\cite{Smith1981} . The discrepancy is less than $\sim 5 \cdot 10
^{-3}$ m$^{-1}$ at all wavelengths, except at 715 nm where the
temperature effect is relevant. Average absorption length for blue
light ($\lambda = 412$ and $\lambda = 440$ nm) is $\sim 50 \div
55$ m; the average attenuation length is $\sim 30$ m. These values
are extremely good when compared to published seawater attenuation
values obtained in conditions of collimated beam and detector
geometry \cite{duntley1963}. The measured blue light attenuation
length value is very close to the ones measured by Khanaev and
Kuleshov \cite{Khanaev1993} in the NESTOR site
\cite{Resvanis1993}.

On the contrary, our results cannot be compared to the ones
published by Bradner et al. \cite{Bradner1981} for the DUMAND
project, Anassontzis et al. \cite{Anassontzis1994} for NESTOR and
the ones measured by the ANTARES collaboration
\cite{Palanque1999}. These measurements were, indeed, carried out
in conditions of not collimated geometry and the measured value is
a quantity usually called {\it effective attenuation coefficient}
$c_{eff}(\lambda)$. This quantity (an {\it apparent optical
property}) is defined as the sum of absorption and only a fraction
of the scattering coefficient:

\begin{equation}
c_{eff}(\lambda)=  a(\lambda) + ( 1-\langle \cos
(\vartheta)\rangle) \cdot b(\lambda)
 \label{eq:effective
attenuation}
\end{equation}

\noindent where $\langle \cos (\vartheta) \rangle$ is the average
cosine of the volume scattering function distribution. This
quantity strongly depends on the amount and dimension of the
dissolved particulate. Measurements carried in ocean
\cite{Mobley1994} gives $\langle \cos (\vartheta) \rangle \sim
0.95$. $c_{eff}(\lambda)$ for NEMO explored sites will be
evaluated only after an {\it in situ} precise measurement of the
volume scattering function scheduled for year 2001.

\section{Acknowledgements}
The NEMO collaboration wants to thank M. Astraldi, G.P. Gasparini
(Istituto di Oceanografia Fisica - CNR, La Spezia) and E.
Accerboni, G. Gelsi, B. Manca, R. Mosetti  (Istituto Nazionale di
Oceanografia e Geofisica Sperimentale, Trieste), M. Leonardi
(Istituto Sperimentale Talassografico - CNR, Messina) C. Viezzoli
(SOPROMAR) for the fruitful collaboration. We want also to thank
Captains V. Lubrano and M. Gentile, officers and crew of the {\it
Urania} Oceanographic Research Vessel, for their outstanding
experience and professionalism shown during the naval campaign.

\end{document}